\documentclass[aps,prl,reprint,amsmath,amssymb,groupaddress,superscriptaddress]{revtex4-1}

\usepackage[utf8]{inputenc}
\usepackage{amsmath}
\usepackage{amsfonts}
\usepackage{amssymb}
\usepackage{bm}
\usepackage{color}
\usepackage{graphicx}
\usepackage{soul}
\usepackage{hyperref}
\usepackage{CJK}
\usepackage{xcolor}
\usepackage{mathtools}
\usepackage{lipsum}

\begin{document}
\begin{CJK*}{UTF8}{gbsn}
\title{{Thermodynamic Bounds on Symmetry Breaking\\in Linear and Catalytic Biochemical Systems}}
\author{Shiling Liang (梁师翎)}
\email{shiling.liang@epfl.ch}
\affiliation{Institute of Physics, School of Basic Sciences, \'Ecole Polytechnique F\'ed\'erale de Lausanne - EPFL, 1015 Lausanne, Switzerland}
\author{Paolo De Los Rios}
\affiliation{Institute of Physics, School of Basic Sciences, \'Ecole Polytechnique F\'ed\'erale de Lausanne - EPFL, 1015 Lausanne, Switzerland}
\affiliation{Institute of Bioengineering, School of Life Sciences, \'Ecole Polytechnique F\'ed\'erale de Lausanne - EPFL, 1015 Lausanne, Switzerland}
\author{Daniel Maria Busiello}
\email{busiello@pks.mpg.de}
\affiliation{Max Planck Institute for the Physics of Complex Systems, 01187 Dresden, Germany}

\begin{abstract}
\noindent
Living systems are maintained out-of-equilibrium by external driving forces. At stationarity, they exhibit emergent selection phenomena that break equilibrium symmetries and originate from the expansion of the accessible chemical space due to non-equilibrium conditions. 
Here, we use the matrix-tree theorem to derive {upper and lower} thermodynamic bounds on these symmetry-breaking features in {linear and catalytic} biochemical systems. Our bounds are independent of the kinetics and hold for both closed and open reaction networks. We also extend our results to master equations in the chemical space. Using our framework, we recover the thermodynamic constraints in kinetic proofreading. Finally, we show that the contrast of reaction-diffusion patterns can be bounded only by the non-equilibrium driving force. Our results provide a general framework for understanding the role of non-equilibrium conditions in shaping the steady-state properties of biochemical systems.
\end{abstract}
\maketitle
\end{CJK*}

\textit{Introduction.}---Living systems operate away from thermodynamic equilibrium, continuously harvesting and consuming energy \cite{fang2019nonequilibrium}. A key feature of out-of-equilibrium living matter is the emergence of various selection phenomena, i.e., {\color{black}biochemical states are populated not only according to their energies, but also to kinetic features} \cite{bertini2004minimum,taniguchi2007onsager,astumian2011stochastic,astumian2019kinetic,busiello2021dissipation,amano2022using}. As a consequence, non-equilibrium settings can break equilibrium energetic symmetries, even if the intimate connection between symmetry breaking and dissipation is still {\color{black}elusive}. Only in recent years, a growing amount of work {\color{black}focused} on highlighting such a link in determined contexts. For example, non-equilibrium conditions have been proven to be crucial to achieve low-error copying mechanisms that preserve DNA and RNA information with an efficiency constrained by the driving force \cite{ehrenberg1980thermodynamic,qian2006reducing,sartori2015thermodynamics,mohr2022individual}. Furthermore, biological systems attain high sensitivity during sensing through non-equilibrium processes \cite{hartich2015nonequilibrium,tu2008nonequilibrium}, and the bi-stability of biochemical states, which is crucial for signaling, can be exploited only away {\color{black}from equilibrium} \cite{vellela2009stochastic}. Also, the newly-discovered mechanism of ultra-affinity in chaperone activity is genuinely out-of-equilibrium and bounded by thermodynamic properties \cite{nguyen2017thermodynamic,de2014hsp70}. On a larger scale, spatial symmetry may be broken by reaction-diffusion patterns, whose emergence is associated with dissipation in the form of energy and information \cite{falasco2018information,nicolis1977self}.

From a broader perspective, the celebrated thermodynamic uncertainty relations set a universal link between dissipation and performance of biological systems, {\color{black}bounding the accuracy of any stochastic current in terms of entropy production} \cite{barato2015thermodynamic,horowitz2020thermodynamic}. However, {\color{black}interesting features are usually} not limited to the statistics of currents, as for most of the cases mentioned above.

A more general, yet not exhaustive, approach to the quest of understanding the role of dissipation in 
biochemical systems has been inspired by network theory. Kirchhoff's law on 
currents has been applied to study thermodynamic constraints in metabolic networks \cite{schilling1999metabolic,schilling2000theory,beard2004thermodynamic,qian2003stoichiometric}, and the representation of stationary states in terms of spanning {\color{black}trees \cite{hill1966studies,schnakenberg1976network} has} been shown to be a powerful tool to derive thermodynamic and structural constraints on non-equilibrium response \cite{owen2020universal,owen2023size}, {\color{black}and connect thermodynamics with kinetic features \cite{polettini2017Effective,ilker2022Shortcuts,khodabandehlou2022Trees,lefebvre2023Frenetic}.}

In this Letter, we give a unified description of thermodynamic bounds on non-equilibrium symmetry breaking in biochemical systems. By taking advantage of the network representation {\color{black}of stationary states}, we derive {thermodynamic} bounds for the ratio of probabilities of any two states (or set of states), which is a measure of selection and, as such, a quantification of the breaking of equilibrium symmetries. We show that these bounds depend solely on network geometry and equilibrium properties, and notably not on the kinetics, as they can be determined by individual spanning trees (for which detailed balance always holds). The derivation of an upper and lower bound {\color{black}(sandwich inequality)} allows us to write down a non-equilibrium phase space, i.e., a region of {\color{black}the chemical space} in which any feasible stationary solution must lie. These results hold for both closed and open {\color{black}catalytic isomerization networks} described in terms of rate equations. In the case of small number of molecules or formation of complexes, our findings can be extended to chemical master equations, providing upper and lower bounds to the ratio of correlation functions of any order {\color{black}for general chemical reaction networks (CRN)}.

{These thermodynamic bounds} can be applied to an immense variety of symmetry-breaking mechanisms in biochemical systems. First, we re-derive in a straightforward way the constraints on kinetic proofreading \cite{ehrenberg1980thermodynamic,qian2006reducing}. Then, we show that the contrast of reaction-diffusion (RD) patterns is limited by the thermodynamic driving force, highlighting a {\color{black}fascinating connection} between observations and non-equilibrium conditions. 


\textit{Results.}---Consider a {reversible} catalytic isomerization network (with uni-molecular {and catalytic reactions)} 
of $N$ species whose probabilities follow a rate equation:
\begin{equation}
    \frac{d}{dt} {p}_i  = \sum_{j(\neq i)} (\hat{k}_{ij} {p}_j - \hat{k}_{ji} {p}_i) \;,
    \label{ME}
\end{equation}
where $p_i$ is the {\color{black}probability of the species $i$, i.e., its concentration,} $c_i$, divided by the total concentration. {\color{black}The transition rate from} $j$ to $i$ might be non-linear, i.e., $\hat{k}_{ij} = \omega_{ij}(\mathbf{p}) k_{ij}$, with $\omega_{ij}(\mathbf{p}) = \omega_{ji}(\mathbf{p})$ and $\mathbf{p}=\{p_1, \dots, p_N\}$, {\color{black}the instantaneous solution of Eq.~\eqref{ME}}. Here, $\omega_{ij}(\mathbf{p})$ encodes {\color{black}catalytic} effects and $k_{ij}$ satisfies {\color{black}local detailed balance, $k_{ij}/k_{ji} = e^{\beta(  F_{ij}-\Delta E_{ij})}$ \cite{maes2021local}.} $\Delta E_{ij} = E_i - E_j$ is the energy difference between states $i$ and $j$, and $F_{ij}$ {\color{black}the non-equilibrium driving force on the edge $(ij)$}. At equilibrium, $F_{ij} = 0$ for all edges, and the stationary solution coincides with the equilibrium Boltzmann distribution, $p_i^{\rm eq} \propto e^{-\beta E_i}$. Out of equilibrium, the steady-state solution, $p^{\rm ss}_i$, depends on non-linear kinetics and topology of the system, not only on state energies. {\color{black}The non-equilibrium driving} can originate from any source, e.g., chemical potential difference \cite{ehrenberg1980thermodynamic,qian2006reducing}, thermal gradient \cite{dass2021equilibrium,liang2022emergent}.

When all the rates are linear, ${p}_i^\mathrm{ss}$ can be written in terms of spanning trees according to matrix-tree theorem \cite{hill1966studies,schnakenberg1976network}. In the non-linear case, {all steady-states satisfying $\omega_{ij}(\mathbf{p}^\mathrm{ss})\neq 0$, $\forall\, i,j$ \footnote{There may exist boundary steady-states with vanishing concentrations that do not satisfy this condition \cite{supplemental_material}. Our derivation does not hold for these solutions.}, can be written through the following spanning-tree decomposition} 
\begin{equation}
    {p}_i^\mathrm{ss} =  \mathcal{N} \sum_\mu A_i(T_\mu;\mathbf{{p}^\mathrm{ss}}) \;,
    \label{SolTree}
\end{equation}
where $\mathcal{N}$ is the normalization factor, and $A_i(T_\mu;\mathbf{p}^\mathrm{ss})$ the product of the rates of the $\mu$-th spanning tree, $T_\mu$, directed towards the state $i$, with the sum running over all possible spanning trees. {\color{black}Since $A_i(T_\mu;\mathbf{p}^\mathrm{ss})$ depends on $\mathbf{p}^{\rm ss}$ in the non-linear case, Eq.~\eqref{SolTree} provides a closed-form solution only for uni-molecular networks, while it must be solved self-consistently in all other cases. Eq.~\eqref{SolTree} can be verified by substituting it into Eq.~\eqref{ME}.}


\begin{figure}[t]
    \centering
    \includegraphics[width=1\columnwidth]{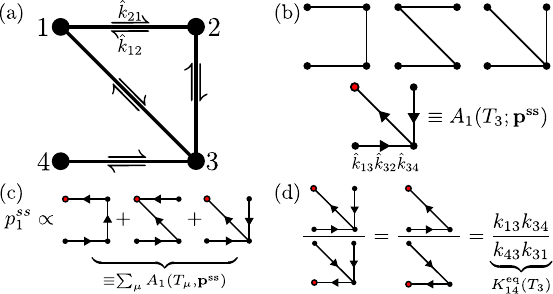}
    \caption{\label{fig:MTT} (a) A four-state reaction network{\color{black}, where edges encode non-linear mechanisms.} (b) All spanning trees, and the directed tree {\color{black}to $1$} constructed from the third spanning tree. (c) Steady-state representation using directed trees. (d) The ratio of two directed trees (to state $1$ and $4$) obtained from the same spanning tree gives a unique reaction pathway {\color{black}without non-linear terms}.}
\end{figure}


To gain intuition about the relevance of spanning trees, consider the ratio $K_{ij}^{\rm eq}(T_\mu)\equiv A_i(T_\mu;\mathbf{p}^\mathrm{ss})/A_j(T_\mu;\mathbf{p}^\mathrm{ss})$. It corresponds to the product of the ratios between forward and backward rates along each edge of the reaction pathway connecting $i$ and $j$ and {\color{black}belonging to $T_\mu$. Indeed, all edges} outside this reaction pathway cancel out in the ratio. Moreover, {\color{black}for each $T_\mu$}, the reaction pathway connecting any two states is unique. We remark that any ratio of forward and backward transition rates along the same edge is $\hat{k}_{lm}/\hat{k}_{ml} = k_{lm}/k_{ml}$ since {\color{black}non-linear terms are symmetric}.
{\color{black}Finally, due to the local detailed balance, $K^{\rm eq}_{ij}(T_\mu)$ is a pseudo-equilibrium quantity,} as it does not depend on the kinetics but only on thermodynamic properties. In Fig.~\ref{fig:MTT}, we {\color{black}show} a simple example.


The main result of this Letter gives an upper and lower bound on the ratio of probabilities of any two {stable steady-states \footnote{It has been conjectured that boundary states are unstable and unreacheable in weakly-reversible CRNs \cite{feinberg1987chemical} and proven in different scenarios \cite{anderson,craciun,autocatalysis,angeli}. The stability requirement is also physically motivated.} \cite{supplemental_material} of a catalytic isomerization CRN}, $i$ and $j$, in terms of pseudo-equilibrium quantities:
\begin{equation}\label{eq:bounds}
\min_{\{T_\mu\}} \left[K^{\rm eq}_{ij}(T_\mu) \right] \leq \frac{p_i^{\rm ss}}{p_j^{\rm ss}}\leq \max_{\{T_\mu\}} \left[K^{\rm eq}_{ij}(T_\mu)\right] \;.
\end{equation}
{\color{black}The derivation relies on the implicit solution in Eq.~\eqref{SolTree}, noting that non-linearities disappear when taking ratios of forward and backward rates \cite{supplemental_material}.} Naively speaking, the maximization (minimization) over all possible spanning trees amounts to maximize (minimize) over all possible reaction pathways connecting $i$ to $j$. The ratio of steady-state probabilities is thus bounded from both sides by the extremal pseudo-equilibrium values it can take, which are sheer thermodynamic quantities independent from non-linearities of reactions and non-equilibrium kinetics \footnote{The derivations of the results presented herein start from the fact that $(\sum_i a_i)/(\sum_i b_i) \leq \max(a_i/b_i)$ if $a_i$ and $b_i$ are positive.}. {\color{black}However, $K^{\rm eq}_{ij}(T_\mu)$ depends on the non-equilibrium driving force along the corresponding pathway from $j$ to $i$ and its logarithm equals the associated entropy production into the environment \cite{peliti2021stochastic}.} As expected, at equilibrium all reaction pathways give the same contribution due to the fact that detailed balance holds, thus upper and lower bounds coincide and are both saturated. This result generalizes similar findings obtained in \cite{cetiner2022reformulating} for a specific case, and in \cite{maes2013heat} only for linear reaction networks. Below, we show that it can be further extended to sets of states, open CRN, and {\color{black}chemical ME}.

\begin{figure}[t]
\includegraphics[width=1\columnwidth]{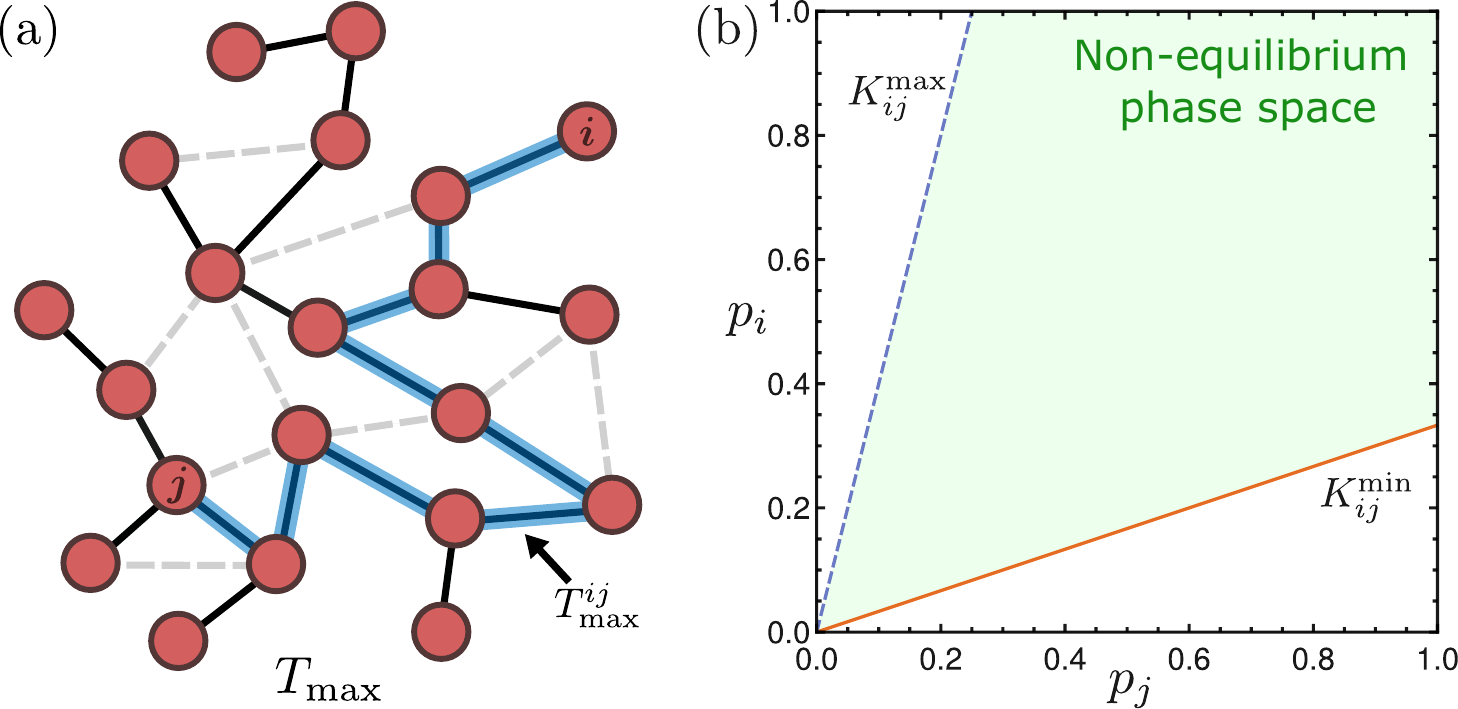}
\caption{\label{fig:chemical_space}(a) In black, {\color{black}the spanning tree providing} the upper bound on $p^{\rm ss}_i/p^{\rm ss}_j$, $T_{\max}$, and its {\color{black}corresponding pathway} $T_{\max}^{ij}$ in blue. Dashed lines represent all other edges of CRN. (b) The non-equilibrium phase space is bounded by pseudo-equilibrium quantities, $K^{\rm max/min}_{ij}$, {\color{black}evaluated from $T_{\rm max/min}$.}}
\end{figure}

In Fig.~\ref{fig:chemical_space}a, we show a pictorial representation of a CRN {\color{black}(where edges can represent non-linear rates), highlighting} the spanning tree determining the upper bound on the symmetry breaking between $p^{\rm ss}_i$ and $p^{\rm ss}_j$, $T_{\max}$.
Fig.~\ref{fig:chemical_space}b presents (in green) the resulting non-equilibrium phase space. At equilibrium, this region collapses into a single line whose slope is determined by the Boltzmann factor $e^{-\beta \Delta E_{ij}}$. Out of equilibrium, we have $K^{\rm eq}_{ij}(T_{\max}) \equiv K^{\rm max}_{ij}$, and analogously we define $K^{\rm min}_{ij}$. 
These bounds determine a region in which steady-state probabilities must lie independently of their kinetic features. Noticeably, for {\color{black}high-dimensional spaces}, the non-equilibrium phase space is a cone whose boundaries are given by bounds on ratios of all pairs of states. Conservation laws, such as mass conservation, shall be introduced as hyperplanes that reduce the accessible region and dimensionality of phase space. {In presence of absolute irreversibility (e.g., unidirectional transitions), a spanning-tree representation cannot be in general obtained \cite{feinberg2019foundations,shinar2010structural}.} 

Eq.~\eqref{eq:bounds} can be readily generalized to evaluate the selection between two groups of states $S = \{s_1, \dots, s_N \}$ and $\Sigma = \{\sigma_1, \dots, \sigma_M \}$, hence quantifying symmetry breaking between coarse-grained chemical macro-states. For the upper bound, we have \cite{supplemental_material}:
\begin{equation}\label{eq:uni_bound}
    \frac{\sum_{i=1}^N p_{s_i}^\mathrm{ss}}{\sum_{i=1}^M p_{\sigma_i}^\mathrm{ss}} \leq \max_{\{T_\mu\}}\left(\frac{\sum_{i=1}^N K^{\rm eq}_{s_i s_N}(T_\mu)}{\sum_{i=1}^M K^{\rm eq}_{\sigma_i \sigma_M}(T_\mu)} K^{\rm eq}_{s_N \sigma_M}(T_\mu) \right),
\end{equation}
where we (arbitrarily) selected $s_N$ and $\sigma_M$ as reference states.
The lower bound can be obtained by maximizing the inverse ratio in the l.h.s. of Eq.~\eqref{eq:uni_bound}.


\textit{Open CRN}.---In the most general case, chemical networks are connected to external reservoirs that keep the concentrations of some chemicals $\bar{\mathbf{c}}_m$ fixed. These couplings lead to {\color{black}additional terms} in the rate equation:
\begin{equation}
    \frac{d}{dt} {c}_i  = \sum_{j(\neq i)} (\hat{k}_{ij} {c}_j - \hat{k}_{ji} {c}_i) + \sum_m (w_{im} \bar{c}_m -w_{mi}{c}_i),
\end{equation}
where $c_i$ indicates the concentrations of the species $i$. 
The spanning tree representation cannot be directly applied, as the chemostatted species are not dynamical variables. However, by merging them into a single moiety $\bar{c}_{\rm e} = \sum_m \bar{c}_m$, {\color{black}the effective} transition rate from $e$ to $i$ through the transition channel $m$ {\color{black}becomes} $w_{im} \bar{c}_m/\bar{c}_{\rm e}$ \cite{supplemental_material}. Using this identification, the resulting rate equation can be solved with the spanning-tree approach {(for steady-states with non-vanishing species concentrations)}, substituting the normalization condition with a constraint on the fixed concentration $\bar{c}_{\rm e}$. As a consequence, the bounds shown above can be equally derived for open {\color{black}catalytic isomerization networks}.

\begin{figure}[t]
\includegraphics[width=\columnwidth]{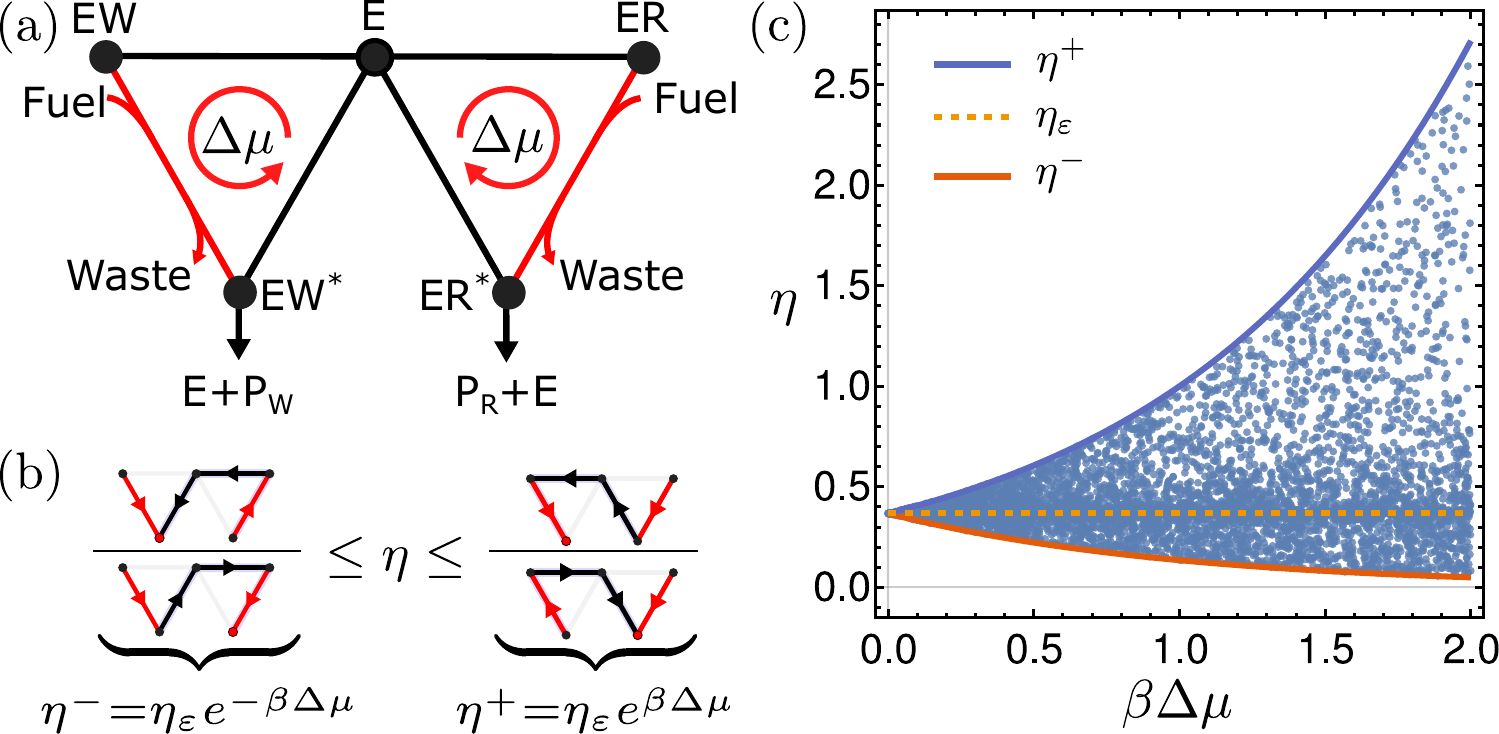}
\caption{\label{fig:proof_reading}(a) Proofreading network driven out of equilibrium by a chemical potential $\Delta\mu$ associated with fuel-to-waste conversion. (b) Spanning trees defining upper and lower bound of the error rate, $\eta\equiv p_\mathrm{EW^*}/p_\mathrm{ER^*}$. (c) Blue points are error rates of the proofreading network with random kinetic parameters but {the same thermodynamic bounds} ({\color{black}red and blue lines}). Dashed-yellow line shows the error rate under equilibrium discrimination.}
\end{figure}

\textit{Kinetic proofreading.}---Living systems encode information in the form of RNA and DNA. 
Genome duplication, translation, and transcription are processes that use this information to select the correct substrate (e.g., nucleotides, codons) with high fidelity, ensuring the survival of the organism \cite{hopfield1974kinetic}. Since the accuracy of this selection is severely limited by equilibrium constraints, back in 1974 Hopfield proposed the existence of energy-consuming intermediate steps that favor kinetic discrimination. The simplest non-equilibrium model for proofreading considers an enzyme, $E$, that can bind wrong ($W$) or right ($R$) substrates, forming complexes. These can be in a passive ($EW$ and $ER$) or active ($EW^*$ and $ER^*$) state, e.g., due to ATP hydrolysis. 
Therefore, the transition from passive to active complexes is driven out-of-equilibrium by a chemical potential $\Delta\mu$ (Fig.~\ref{fig:proof_reading}a). 
It has been shown that the energetic cost of performing a cycle \cite{ehrenberg1980thermodynamic,qian2006reducing} constrains the proofreading performance, i.e., $\eta \equiv p_\mathrm{EW^*}/p_\mathrm{ER^*} \geq \exp\left(-\beta(\epsilon+\Delta\mu)\right)$. In its original formulation, this bound results from an optimization {\color{black}on} reaction rates. Here, by employing {the thermodynamic bounds}, we have:
\begin{equation}
    \underbrace{\min_{\{T_\mu\}}\left(\frac{{A}_\mathrm{EW^*}(T_\mu)}{{A}_\mathrm{ER^*}(T_\mu)}\right)}_{\eta^-=\eta_\varepsilon e^{-\beta\Delta\mu}}
    \leq
    \eta\leq \underbrace{\max_{\{T_\mu\}}\left(\frac{{A}_\mathrm{EW^*}(T_\mu)}{{A}_\mathrm{ER^*}(T_\mu)}\right)}_{\eta^+=\eta_\varepsilon e^{\beta\Delta\mu}},
\end{equation}
where $\eta_\varepsilon=e^{-\beta\varepsilon}$ is the equilibrium error rate due to energetic discrimination 
(see Fig.~\ref{fig:proof_reading}b). The driving force can either increase or suppress the error rate, depending on the kinetic properties of the network, as shown in Fig.~\ref{fig:proof_reading}c. Our approach recovers the infinite-energy Hopfield rate and can be readily used to find thermodynamic bounds on multi-stage proofreading \cite{supplemental_material}, a process with too many parameters to be bounded via optimization.

\textit{RD patterns.}---Originally illustrated in the seminal work of Turing \cite{turing1952chemical}, the instability of the homogeneous fixed point in reaction-diffusion systems, with the consequent emergence of spatial patterns, {\color{black}attracted huge attention} \cite{biancalani2010stochastic,nakao2010turing,asllani2014turing}. RD patterns have also been analyzed from a thermodynamic perspective and their onset classified as a non-equilibrium phase transition \cite{falasco2018information}. Recently, Brauns et al. \cite{brauns2020phase} proposed a phase-space geometric approach to study pattern formation of mass-conserving reaction-diffusion systems. Consider the following dynamics for concentrations $u$ and $v$:
\begin{eqnarray}\label{eq:rd-equation}
    \partial_t u &=& D_u\nabla^2 u + f(u,v), \nonumber \\
    \partial_t v &=& D_v\nabla^2 v - f(u,v) \;.
\end{eqnarray}
The reaction term $f(u,v)$ contains all the transitions connecting $u$ and $v$, i.e., $f(u,v)\equiv \sum_i (\hat{k}^{(i)}_{uv}v-\hat{k}_{vu}^{(i)}u)$, 
with $(i)$ indicating the reaction channel, while $D_u$ and $D_v$ are diffusion coefficients. In \cite{brauns2020phase}, it is shown that all the values of concentrations explored by a spatially extended pattern must be embedded between the intersection of the \textit{reactive nullcline}, $f(u,v) = 0$, and the \textit{flux-balance subspace}, $D_u u + D_v v = \eta_0$, where $\eta_0$ is a constant determined by the \textit{total turnover balance}. 
{\color{black}Since the \textit{reactive nullcline} is determined by steady-state solutions, it must be embedded inside the non-equilibrium phase space.} 
Here, we study which limits to RD patterns may be set only according to the laws of thermodynamics.

\begin{figure}[t]
    \includegraphics[width=1\columnwidth]{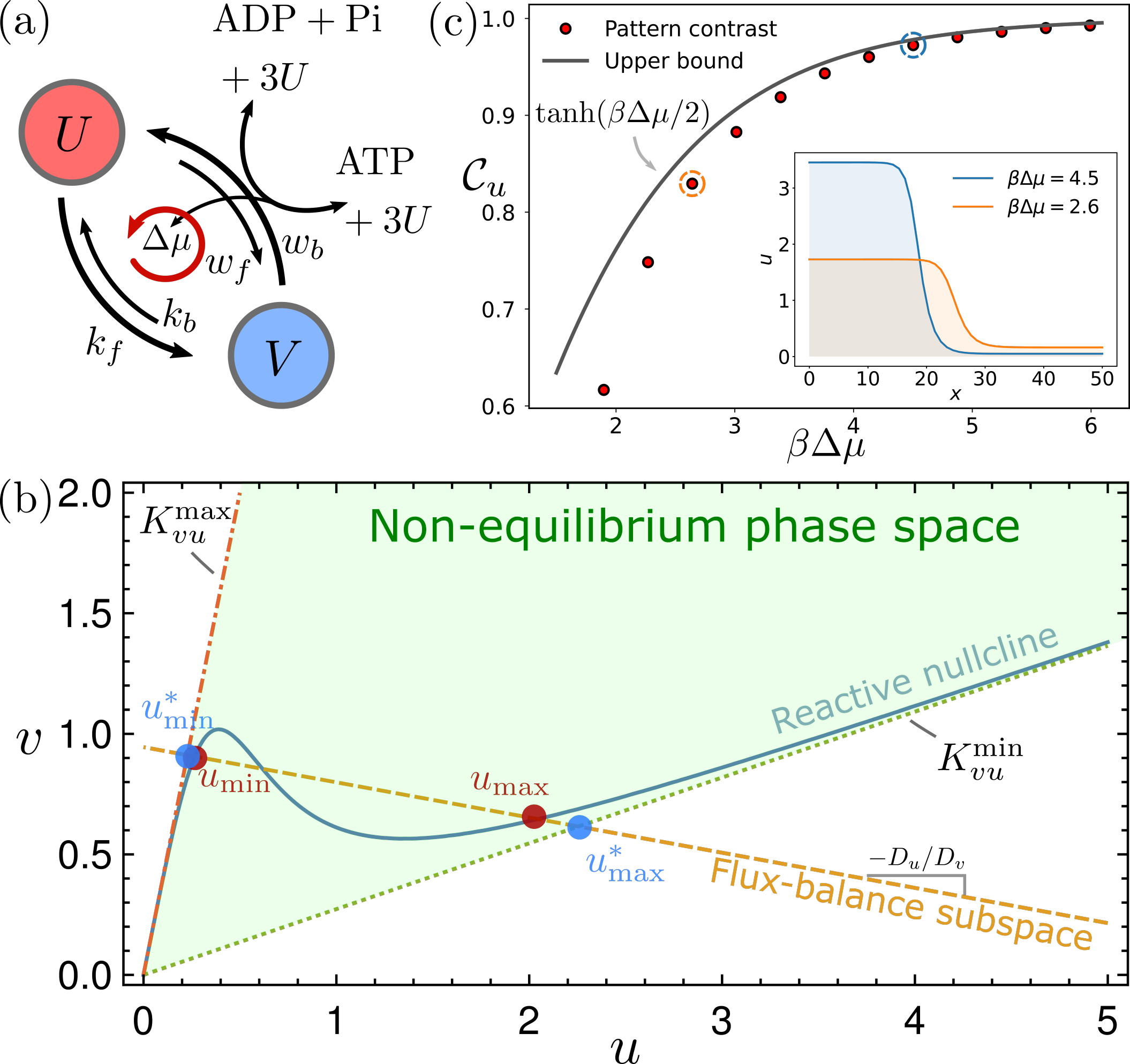}
    \caption{Thermodynamic bound on the contrast of RD patterns. (a) Model system under investigation, driven out of equilibrium by ATP hydrolysis. (b) {Thermodynamic bounds} (in blue) determine the non-equilibrium phase space and contain the actual range of the observed pattern (in red). (c) Pattern visibility is upper bounded by Eq.~\eqref{constrast}. {\color{black}In subpanel, we plot patterns corresponding to coloured circles.}}
    \label{fig:RD pattern}
\end{figure}

In Fig.~\ref{fig:RD pattern}a, we show the system under investigation. Fig.~\ref{fig:RD pattern}b shows, in red, the maximum and minimum accessible values of $u$ and $v$ in the pattern. {The thermodynamic bounds} {\color{black}on $(u/v)$} at stationarity determine the non-equilibrium phase space to which the concentrations must belong. The intersections between these bounds and the flux-balance subspace define the extremal accessible values for $u$, $u^*_{\rm max/min}$ (see Fig.~\ref{fig:RD pattern}b). {From \cite{supplemental_material}}:
\begin{equation}
    \frac{u_\mathrm{max}}{u_\mathrm{min}}\leq \frac{u_\mathrm{max}^*}{u_\mathrm{min}^*}\leq e^{\beta\Delta\mu},
    \label{RDbounds}
\end{equation}
where $\beta \Delta\mu$ is the non-equilibrium driving force provided by ATP hydrolysis. The rightmost equality is reached when {\color{black}the system is dominated by the non-linear pathway, while the leftmost equality is attained when $D_u/D_v \to 0$, i.e. uniform concentration of $v$.}


At this point, we define the pattern contrast (or visibility) as $\mathcal{C}_x = (x_{\rm max} - x_{\rm min})/(x_{\rm max} + x_{\rm min})$, with $x = u, v$ 
From Eq.~\eqref{RDbounds}, we readily obtain \footnote{Let $\alpha = u_\mathrm{max}/u_\mathrm{min}$, then $\mathcal{C}_u = \frac{\alpha-1}{\alpha+1}=1-\frac{2}{\alpha+1}\leq1-\frac{2}{e^{\beta\Delta\mu}+1}=\tanh(\beta\Delta\mu/2)$}:
\begin{equation}
    \mathcal{C}_u \leq \tanh\left(\frac{\beta \Delta\mu}{2}\right) \;.
    \label{constrast}
\end{equation}
This inequality immediately tells that pattern formation is a sheer consequence of non-equilibrium conditions, as at equilibrium the {\color{black}accessible} space shrinks into a line and $\mathcal{C}_u$ goes to zero. Moreover, we only need to know the thermodynamic force driving the system out of equilibrium to bound the contrast of a pattern, independently of the details of the kinetics. In Fig.~\ref{fig:RD pattern}c, we numerically check Eq.~\eqref{constrast} in our example. A tighter thermo-kinetic bound for $\mathcal{C}_u$ can be found by adding information about diffusion coefficients \cite{supplemental_material}. Although we inspected a simple two-state system, the presented approach is valid for any mass-conserving RD system.

\textit{Chemical ME.}---CRNs with small number of molecules or formation of complexes are impossible to describe using Eq.~\eqref{ME}. Nevertheless, it is always possible to write down a chemical {\color{black}ME}, where each state is $\mathbf{n} = \{ n_1, \dots, n_N\}$, i.e., the number of molecules of each of the $N$ species. {\color{black}Its} solution can be found using the spanning-tree method in the entire chemical space. Employing the same methodology leading to Eqs.~\eqref{eq:bounds} and \eqref{eq:uni_bound}, we can bound the ratio between correlation functions of any order \cite{supplemental_material}. For $l$-th order correlations, we have:
\begin{equation}
    \frac{\langle n_{i_1} \dots n_{i_l} \rangle^{\rm ss}}{\langle n_{j_1} \dots n_{j_l} \rangle^{\rm ss}} \leq \max_{\{T_\mu\}} \left( \frac{\sum_{\mathbf{n}} \mathbf{a}_l ~K^{\rm eq}_{(\mathbf{i}_l,\mathbf{a}_l),(\mathbf{i}_l,\mathbf{L})}}{\sum_{\mathbf{n}} \mathbf{a}_l ~K^{\rm eq}_{(\mathbf{j}_l,\mathbf{a}_l),(\mathbf{j}_l,\mathbf{L})}} K^{\rm eq}_{(\mathbf{i}_l,\mathbf{L}),(\mathbf{j}_l,\mathbf{L})} \right),
    \label{chembound}
\end{equation}
where the dependence on $T_\mu$ has been omitted for clarity. Here, $\mathbf{i}_l = (i_1, \dots, i_l)$, $\mathbf{a}_l = (a_1, \dots, a_l)$, and $K^{\rm eq}_{(\mathbf{i}_l,\mathbf{a}_l),(\mathbf{i}_l,\mathbf{L})}$ is evaluated from the pathway connecting the states in which $\mathbf{i}_l = \mathbf{a_l}$ and $\mathbf{i}_l = \mathbf{L}$, an (arbitrary) reference state. In \cite{supplemental_material}, we show that Eq.~\eqref{chembound} can be upper bounded by one single pathway. When $l=1$, Eq.~\eqref{chembound} bounds the ratio of average concentrations, as in Eq.~\eqref{eq:bounds}. When $l=2$, the thermodynamic bound applies to ratio of correlations and might be useful to study how couplings inferred from correlation matrix (e.g., using a maximum-entropy approach \cite{bialek2012statistical,castellana2014inverse,de2018introduction}) depend on non-equilibrium conditions. Eq.~\eqref{chembound} is manifestly similar to Eq.~\eqref{eq:uni_bound} and the lower bound can be obtained from {\color{black}the inverse ratio}. A detailed analysis of applications and criticalities of {\color{black}these {thermodynamic} bounds for Chemical ME is left for future studies}. 




\textit{Discussion.}---In this Letter, we {derive upper and lower thermodynamic} bounds on quantities that {\color{black}assess} how much equilibrium symmetries are broken by non-equilibrium drivings. We presented two simple systems in which our framework provide interesting results, in particular highlighting the bound on pattern contrast solely as a function of the thermodynamic force. The general advantage of our findings is that they are independent of non-linear reaction terms and kinetics, usually very important to understand systems operating away from equilibrium. As such, unraveling a reformulation of these bounds in the contexts of information thermodynamics might pave the way to understand {\color{black}thermodynamic} constraints on chemical information-processing devices \cite{amano2022insights,flatt2021abc}.

{\color{black}Countless examples} of selection phenomena have proven to be relevant in as many different contexts. Surely enough, the generality of our results will be a powerful tool to understand potentialities and limitations of non-equilibrium conditions in shaping the steady-state properties of biochemical systems. Indeed, nonequilibrium drivings expand the accessible chemical space, enabling living systems to adapt to various {\color{black}non-linear} mechanisms and exhibit complex behaviours. For example, the emergence of chiral symmetry breaking in thermodynamically consistent {\color{black}CRN} is a {\color{black}intriguing} open problem in which the role of {\color{black}non-equilibrium conditions} has not been thoroughly explored yet \cite{blackmond2009if}. {\color{black}On theoretical grounds, our results might contribute to the promising research on sandwich inequalities \cite{maes2013heat,liang2023thermodynamic,manzano2023thermodynamics}.}

{\color{black}Finally, the approach employed for catalytic isomerization networks might be extended to deal with CRNs represented by hypergraphs \cite{dalcengio2023Geometry}, underscoring the broader relevance of an accessible chemical space shaped by non-equilibrium drivings.}

\section{Acknowledgments}

\begin{acknowledgments}
S.L. appreciates helpful discussions with Hong Qian and the suggestion of the visualized proof of inequality in Supplemental Material from Xiaochen Fu. S.L. and P.D.L.R. thank the Swiss National Science Foundation for support under grant 200020\_178763.
\end{acknowledgments}

\bibliography{refs}

\end{document}


\begin{CJK*}{UTF8}{gbsn}
\title{Supplemental Material for: \\ ``{Thermodynamic Bounds on Symmetry Breaking\\ in Linear and Catalytic Biochemical Systems}''}
\author{Shiling Liang (梁师翎)}
\affiliation{Institute of Physics, School of Basic Sciences, \'Ecole Polytechnique F\'ed\'erale de Lausanne - EPFL, 1015 Lausanne, Switzerland}
\author{Paolo De Los Rios}
\affiliation{Institute of Physics, School of Basic Sciences, \'Ecole Polytechnique F\'ed\'erale de Lausanne - EPFL, 1015 Lausanne, Switzerland}
\affiliation{Institute of Bioengineering, School of Life Sciences, \'Ecole Polytechnique F\'ed\'erale de Lausanne - EPFL, 1015 Lausanne, Switzerland}
\author{Daniel Maria Busiello}
\affiliation{Max Planck Institute for the Physics of Complex Systems, 01187 Dresden, Germany}
\begin{abstract}

\end{abstract}
\maketitle
\end{CJK*}
\onecolumngrid
\section{Derivation of the Universal Thermodynamic Bounds}
\subsection{On ratios of steady-state probabilities}

We consider {reversible} chemical reaction networks (CRNs) in which catalytic, auto-catalytic, and isomerization reactions can occur, i.e. catalytic isomerization networks. They constitute a broad class of reaction schemes that describe a multitude of biochemical systems. Reactions between any pair of chemicals are as follows
\begin{equation}
    \sum_{m}\nu_m X_m + X_i\xrightleftharpoons[k_{ij}]{k_{ji}}X_j +\sum_{m}\nu_m X_m
\end{equation}
where $\sum_{m}\nu_m X_m$ represents the catalytic mechanism and $\nu_m$ is the number of species $X_m$ involved in the reaction as catalysts. Such a reaction network follows mass action kinetics and can be highly nonlinear. However, it can still be represented by a graph since the non-linear dependency given by catalytic species can be absorbed into the transition rates (edges of the graph). This procedure leads to the following (general) form of reaction:
\begin{equation}
     X_i\xrightleftharpoons[k_{ij} \prod_{m}{[X_i]}^{\nu_m } ]{k_{ji}\prod_{m}[X_i]^{\nu_m}}X_j \;,
\end{equation}
where we introduce the effective nonlinear rate $\hat{k}_{ij}\equiv k_{ij}\prod_m[X_m]^{\nu_m}$, with $[X_m]$ is the concentration of chemical species $X_m$. This rate becomes linear when catalysts are not present. Rewriting concentrations as probability distributions by adding a proper normalization, i.e., $p_m = [X_m]/\sum_m[X_m]$, we get the following nonlinear rate equation shown in the main text:
\begin{equation}
\frac{dp_i}{dt} = \sum_{j(\neq i)} \hat{k}_{ij}p_j-\hat{k}_{ji}p_i,
\end{equation}
with each pair of transition rates satisfying the local detailed balance relation
\begin{equation}\label{eq:local detailed balance}
\frac{\hat{k}_{ij}}{\hat{k}_{ji}}=\frac{k_{ij}}{k_{ji}} = e^{\beta (F_{ij}-\Delta E_{ij})}\;,
\end{equation}
with $F_{ij}$ the non-equilibrium thermodynamic force from $i$ to $j$. The non-linearity of the rates is due to catalytic reactions and encoded in the prefactor $\omega_{ij}$, i.e., $\hat{k}_{ij}=\omega_{ij}(\mathbf{p})k_{ij}$ and $\hat{k}_{ji}=\omega_{ji}(\mathbf{p})k_{ji}$. These non-linear factors have to be the same for forward and backward reaction, $\omega_{ij}(\mathbf{p})=\omega_{ji}(\mathbf{p})$, so that the local detailed balance relation, Eq.~\eqref{eq:local detailed balance}, is guaranteed. This condition stems from the fact that every chemical reaction must be reversible and ensures thermodynamic consistency.


The stationary solution of the non-linear reaction network cannot be explicitly obtained in general. However, the final stationary state have to satisfy, by definition, the following equation:
\begin{equation}\label{eq:nonlinear_master}
    \forall i: 0 =\sum_{j(\neq i)} \hat{k}_{ij}p_j^\mathrm{ss}-\hat{k}_{ji}p_j^\mathrm{ss}
\end{equation}
{When the stationary state further satisfies $\omega_{ij}(\mathbf{p}^{\mathrm{ss}})\neq 0$, $\forall\,i,j$, Eq.\eqref{eq:nonlinear_master}} can be rewritten in terms of the spanning trees of the reaction network as follows: 
\begin{equation}
p_k^\mathrm{ss}=\frac{\sum_\mu{A}_k(T_\mu;\mathbf{p^\mathrm{ss}})}{\sum_k\sum_\mu{A}_k(T_\mu;\mathbf{p^\mathrm{ss}})}
\label{eq:matrixtree}
\end{equation}
Notice that, in the linear case, i.e., without (auto-)catalytic reactions, the r.h.s. of this equation does not depend on the state of the system and it provides a closed solution for the steady-state probabilities. In a more general non-linear scenario, the r.h.s. does depend on $\mathbf{p}^{\rm ss}$ itself and the equation has to be solved self-consistently, thus being not particularly helpful to find steady-state probabilities. However, formally Eq.~\eqref{eq:matrixtree} is valid in all cases {with an invertible transition matrix} (whether as a closed or self-consistent equation). This can be easily seen by substitution and comes from the fact that, once we know $\mathbf{p}^{\rm ss}$, inserting this solution back into the rates, they effectively become numbers and the steady-state can be found by standard techniques for linear master equations. {In all cases in which the transition matrix is not invertible due to the fact that $\omega_{ij}(\mathbf{p}^\mathrm{ss})=0$ for a set of edges, the corresponding solution does not satisfy Eq.~\eqref{eq:matrixtree}. These steady states that are characterized by one (or more) vanishing concentration are also named boundary states \cite{feinberg2019foundations} and our framework cannot be applied to them.
Let us start with a simple example on this scenario, i.e., the autocatalytic reaction 
\begin{equation}
    A + B\xrightleftharpoons[k_b]{k_f} 2A
\end{equation}
that admits two solutions. The stable solution is $p_A^\mathrm{ss} = 1/(1+k_b/k_f)$ which satisfies the spanning tree decomposition shown above. The other solution is $p_A=0$ which is obtained noting that $\hat{k}_f=k_f[A]^\mathrm{ss}=0$ and $\hat{k}_b = k_b [A]^\mathrm{ss}=0$. In this simple system, it is easy to show that this second solution is unstable. 

It seems intuitively reasonable that all boundary states are unstable in reversible CRNs, as whenever a concentration is slightly perturbed, the transition $\omega_{ij}(\mathbf{p}^\mathrm{ss})$ that was zero becomes non-zero. As a consequence, the matrix is invertible and, as such, solvable through the spanning tree decomposition which does not admit steady states with vanishing concentrations. This reasoning is, of course, not a proof. Indeed, for general scenarios, it has been conjectured that boundary states are unstable and unreachable in weakly-reversible CRNs \cite{feinberg1987chemical,feinberg2019foundations}. Despite proofs have been proposed in several contexts \cite{anderson,craciun,autocatalysis,angeli}, a general demonstration of this conjecture is still lacking. Apart from having only non-zero concentrations, it is physically plausible, more in general, to restrict our formulation to stable steady-states when dealing with biophysical and biochemical applications.

Here, we do not want to delve into this discussion about the stability of steady states of reversible CRNs, but we emphasize that our framework holds for all non-boundary (possibly stable) steady states of reversible CRNs composed by isomerization and catalytic reactions.}

\begin{figure}[t]
\centering
\includegraphics[width=.8\columnwidth]{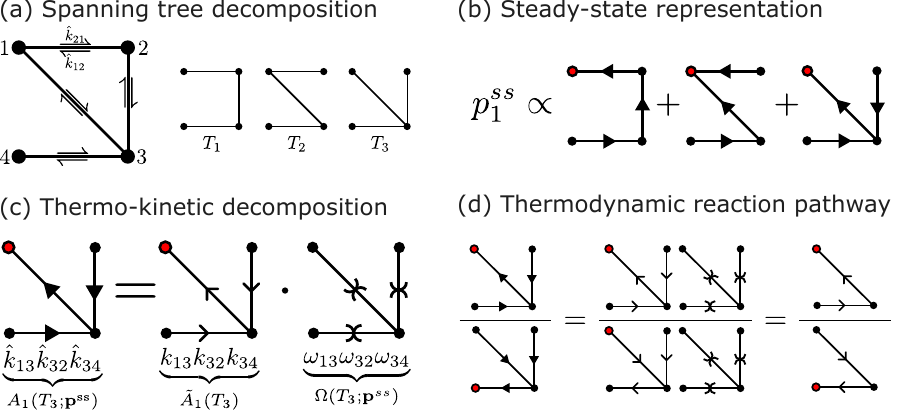}
\caption{\label{fig:S1}(a) Spanning tree decomposition for a simple network. (b) Contribution to $p_1^{\rm ss}$, i.e., numerator of the r.h.s. of Eq.~\eqref{eq:matrixtree}. (c) Decomposition into thermodynamic and kinetic part of a specific tree. (d) Proof of the fact that any ratio between $A_i$ and $A_j$ does not depend on the kinetic term.}
\end{figure}

In Fig.~\ref{fig:S1}(a-b), we show the spanning tree decomposition in a particularly simple case of a four-state network (also presented in Fig.~1 of the main text). Fig.~\ref{fig:S1}(c) shows the decomposition between thermodynamic and kinetic (non-linear) parts, while Fig.~\ref{fig:S1}(d) highlights that the symmetric features of $\omega_{ij}$ makes the ratio of $A_i$ and $A_j$, for any $i$ and $j$, independent from $\mathbf{p}^{\rm ss}$.

\begin{figure}[h]
\centering
\includegraphics[width=.4\columnwidth]{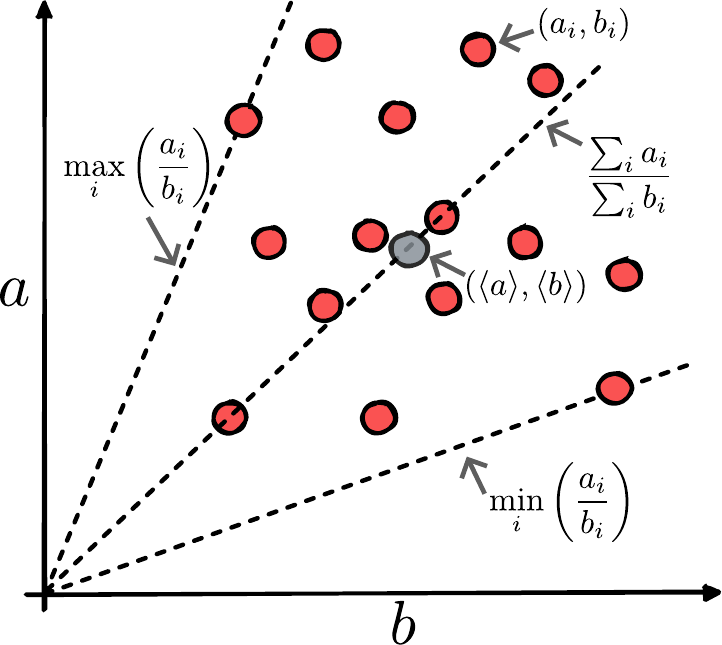}
\caption{\label{fig:S2}Visualization of the inequality in Eq.~\eqref{eq:inequality}, i.e., $\min_i(a_i/b_i)\leq\sum_ia_i/\sum_ib_i\leq\max_i(a_i/b_i)$, which is used to prove the universal thermodynamic bounds presented in the main text.}
\end{figure}

We can make use of Eq.~\eqref{eq:matrixtree} to express any ratio of probabilities:
\begin{equation}
    \frac{p_i^\mathrm{ss}}{p_j^\mathrm{ss}} = \frac{\sum_\mu{A}_i(T_\mu;\mathbf{p^\mathrm{ss}})}{\sum_\mu{A}_j(T_\mu;\mathbf{p^\mathrm{ss}})}
    \label{eq:boundpss}
\end{equation}
This equation is still self-consistent, as it depends on the stationary solution on both sides. Noting that the following mathematical inequalities always hold (see also Fig.~\ref{fig:S2}):
\begin{equation}
     \frac{\sum_i a_i}{\sum_i b_i} = \frac{\sum_i b_i (a_i/b_i)}{\sum_i b_i} \in\left[\min_i \left( \frac{a_i}{b_i} \right), \max_i \left( \frac{a_i}{b_i} \right)\right] \qquad a_i,b_i>0 \;.
    \label{eq:inequality}
\end{equation}
Its derivation goes as follows. Letting $K^{\max} = \max_{i}(a_i/b_i)$, then by definition we have $b_i K^{\max}\geq a_i$. Summing over the index $i$ we get $\sum_i(K^{\max}b_i)=K^{\max}\sum_ib_i\geq\sum_ia_i$. By divingding both side by $\sum_i b_i$ and writting down the definition of $K^{\max}$ explicitely, we get the desired inequality, $\sum_i b_i/\sum_i a_i\geq \max_i(a_i/b_i)$. The equal sign is saturated when $a_i/b_i$ for any $i$ takes the same value, thus upper and lower bounds coincide and are both saturated. The bounds presented in the main text immediately follow by applying Eq.~\eqref{eq:inequality} to Eq.~\eqref{eq:boundpss}. For the upper bound, we have:
\begin{equation*}
    \frac{p^{\rm ss}_i}{p^{\rm ss}_j} \leq \max_{T_\mu} \frac{A_i(T_\mu;\mathbf{p}^{\rm ss})}{A_j(T_\mu;\mathbf{p}^{\rm ss})} = \frac{A_i(T_{\max};\mathbf{p}^{\rm ss})}{A_j(T_{\max};\mathbf{p}^{\rm ss})} = \prod_{(m,n) \in T^{(ij)}_{\max}} \frac{\hat{k}_{mn}}{\hat{k}_{nm}} = \prod_{(m,n) \in T^{(ij)}_{\max}} \frac{k_{mn}}{k_{nm}} =\prod_{(m,n) \in T^{(ij)}_{\max}}e^{\beta ( F_{mn}-\Delta E_{mn} )} = K^{\rm eq}_{ij}(T_{\max})
\end{equation*}
where we used that the only surviving terms in the ratio are those belonging to the pathway connecting $i$ with $j$, and that the sum of energy differences and forces in the exponents gives the total energy difference and the total force between $i$ and $j$. The same strategy can be applied to obtain the lower bound. Indeed, as explained in the main text and shown above, taking the ratio between the forward and backward product of rates along the same pathway, non-linearities cancel out (see also Fig.~\ref{fig:S1}(d)).

Notice that $K^{\rm eq}_{ij}(T_{\max})$ is a pseudo-equilibrium quantity as it comes from the local detailed balance. Moreover, it quantified the value of $p^{\rm ss}_i/p^{\rm ss}_j$, if the system were connected - in this case - only through $T_{\max}$.

\subsection{On ratios of coarse-grained steady-state probabilities}
{From now on, our framework is used without explicitly mentioning all the aforementioned necessary requirements to build an effective graph for catalytic isomerization chemical networks.}

Sometimes we only have access to the concentrations of coarse-grained chemical states, i.e., a set of states impossible to resolve from each other. The universal thermodynamic bounds on the selection parameter can be obtained even in this case. To this aim, consider two sets of states, $S$ and $\Sigma$. We arbitrarily choose a reference state, say $s_N$ for $S$ and $\sigma_M$ for $\Sigma$. Then, we compute all the ratios within each set with respect to this state:
\begin{equation}
    \frac{\sum_{s_k\in S}p_{s_k}^\mathrm{ss}}{\sum_{\sigma_k\in \Sigma}p_{\sigma_k}^\mathrm{ss}}
    =\frac{\sum_{s_k\in S}\sum_\mu{A}_{s_k}(T_\mu;\mathbf{p^\mathrm{ss}})}{\sum_{\sigma_k\in \Sigma}\sum_\mu{A}_{\sigma_k}(T_\mu;\mathbf{p^\mathrm{ss}})}
    =\frac{\sum_\mu A_{s_N} \left(\sum_{s_k\in S}{A}_{s_k}/A_{s_N}\right)}{\sum_\mu A_{\sigma_N} \left(\sum_{\sigma_k\in \Sigma}{A}_{\sigma_k}/A_{\sigma_N}\right)}
    \label{multistates}
\end{equation}
Noting again that all ratios do not depend on steady probabilities, we can apply the inequality in Eq.~\eqref{eq:inequality} to Eq.~\eqref{multistates} and readily obtain the following upper bound:
\begin{equation}
    \frac{\sum_\mu A_{s_N} \left(\sum_{s_k\in S}{A}_{s_k}/A_{s_N}\right)}{\sum_\mu A_{\sigma_N} \left(\sum_{\sigma_k\in \Sigma}{A}_{\sigma_k}/A_{\sigma_N}\right)} \leq \max_{\{T_\mu \}} \left( \frac{\sum_{i=1}^N K^{\rm eq}_{s_i s_N}}{\sum_{i=1}^N K^{\rm eq}_{\sigma_i \sigma_N}} K^{\rm eq}_{s_N \sigma_N} \right) \;.
\end{equation}
The lower bound can be analogously derived. Despite a more complex structure with respect to the previous case, these bounds again depend solely on the thermodynamic forces acting on the underlying reaction networks.

The validity of these slightly more complex bounds stems directly from the one of the universal thermodynamic bounds for single species (see the section above).

\section{Bounds for open chemical reaction networks}
\begin{figure}[h]
\centering
\includegraphics[width=1\columnwidth]{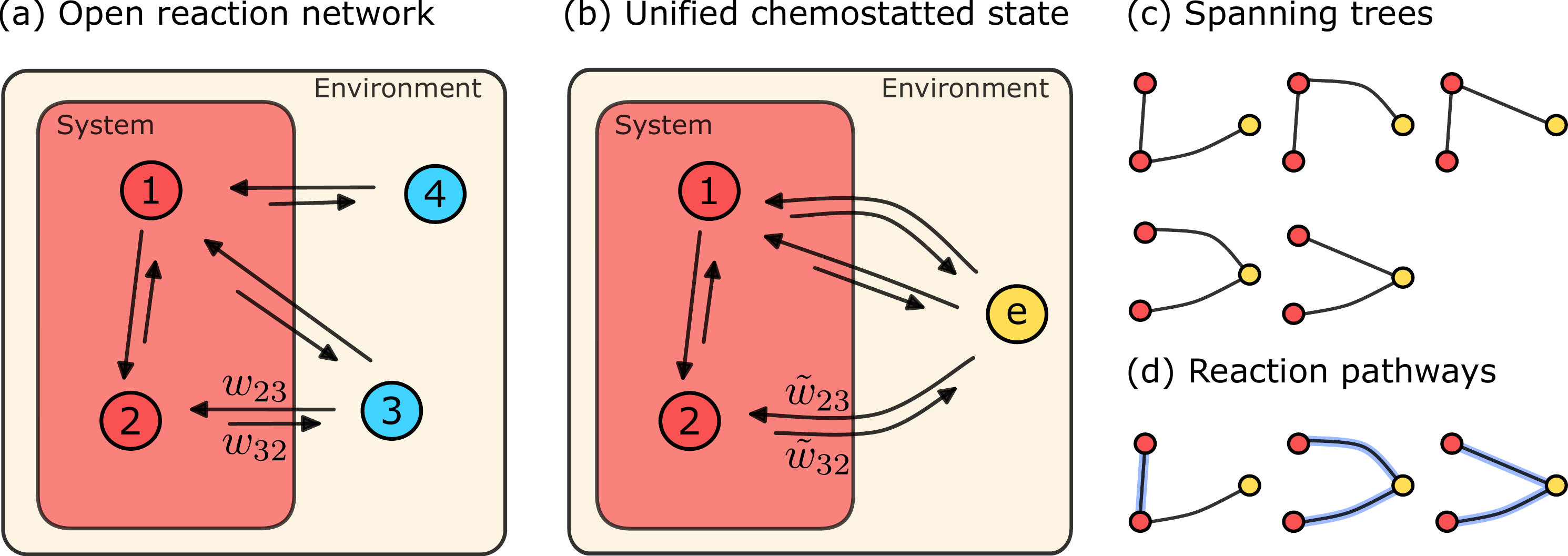}
\caption{\label{fig:S3}Network decomposition of an open chemical reaction network. (a) An open chemical reaction network with two chemostatted species. (b) The two chemostatted species can be merged into a single moiety with concentration $\bar{c}_e = \bar{c}_3 + \bar{c}_4$. (c) Spanning tree decomposition of the open chemical reaction network. (d) The reaction pathways between state 1 and state 2 are identified from the spanning trees.}
\end{figure}

Here, we present the derivation of the bound in the case of open CRNs. The rate equation for these systems is:
\begin{equation}
    \frac{dc_i}{dt}=\sum_{j(\neq i)}(\hat{k}_{ij}c_j-\hat{k}_{ji}c_i)+\sum_m(w_{im}\bar{c}_m-w_{mi}c_i)
\end{equation}
Here, $\bar{c}_m$ are the chemostatted species whose concentrations are kept constant by external reservoirs. In Fig.~\ref{fig:S3}(a), we present a simple example of this setting. Of course, the spanning tree method cannot be straightforwardly applied, as $\bar{c}_m$ are not dynamical variables, being fixed in time by external constraints.

As graphically shown in Fig.~\ref{fig:S3}(b), we can merge all $\tilde{c}_m$ into a fictitious chemical species of concentration $\bar{c}_{\rm e} = \sum_m \bar{c}_m$. The resulting rate equation in terms of this reduced set of states is:
\begin{equation}
    \sum_{j(\neq i)}(\hat{k}_{ij}c_j-\hat{k}_{ji}c_i)+\sum_m \Big( \underbrace{\frac{w_{im}\bar{c}_m}{\sum_m \bar{c}_m}}_{\tilde{w}_{im}}\underbrace{\sum_m \bar{c}_m}_{c_{\rm e}} -\underbrace{w_{mi}}_{\tilde{w}_{mi}}c_i \Big)
    =\sum_{j(\neq i)}(\hat{k}_{ij}c_j-\hat{k}_{ji}c_i)+\sum_m (\tilde{w}_{im}\bar{c}_{\rm e}-\tilde{w}_{mi}c_i) \;.
    \label{eq:openME}
\end{equation}
Now, each $m$ identifies a different reaction channel and the spanning tree method can be applied just by changing the normalization condition. Notice that, instead of fixing one of the concentrations so that the sum of them will be $c_{\rm tot}$, the total concentration, as for closed CRNs, here we have to fix $\bar{c}_e$ to be equal the value externally imposed by chemostats. An additional caveat is that each transition channel has to be considered independently when constructing all possible spanning trees. Apart from these two modifications, the universal thermodynamic bounds can be derived following the same steps explained above, again starting from the self-consistent solution of the non-linear rate equations resulting from Eq.~\eqref{eq:openME} where $\bar{c}_e$ acts as a fictitious variables. In Fig.~\ref{fig:S3}(c-d), we show how to identify trees and reaction pathways. The fact that non-linearities cancel out when considering ratio of forward and backward rates stems from the symmetry of catalytic terms, as already discussed.

\section{Kinetic proofreading}

\begin{figure}[t]
\centering
\includegraphics[width=.8\columnwidth]{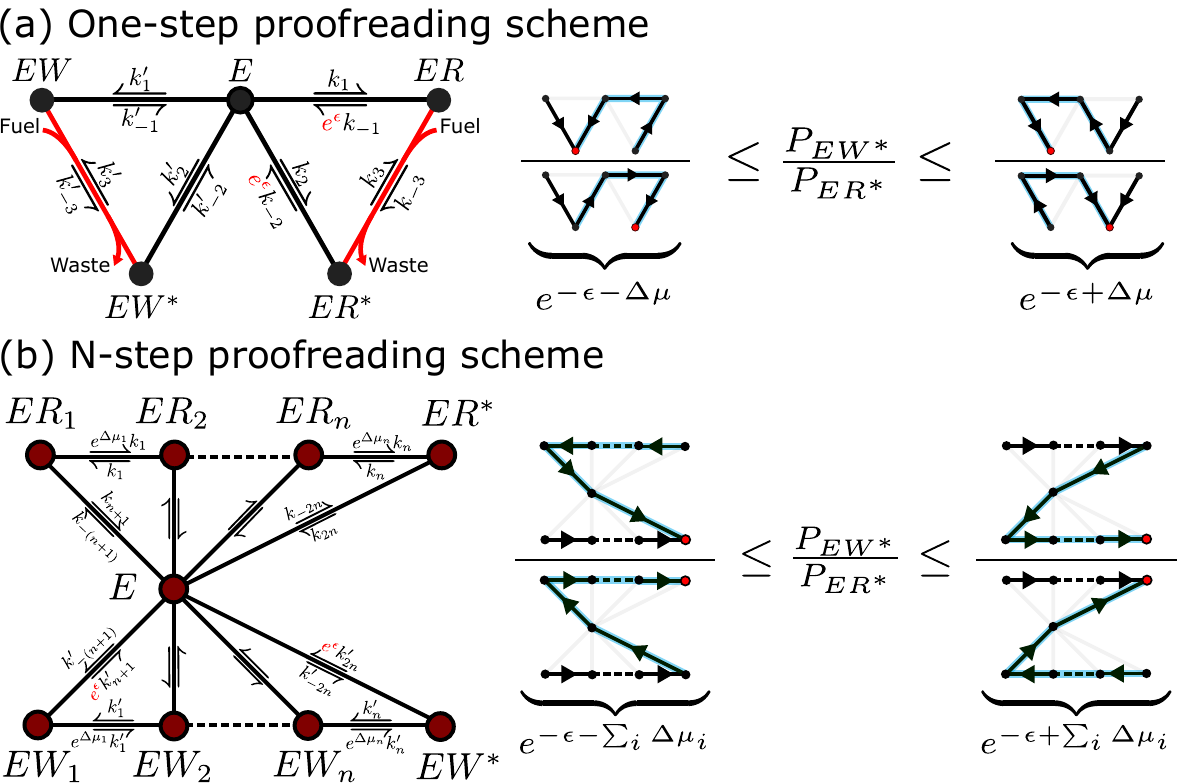}
\caption{\label{fig:proofreading_SI}Proofreading networks and the corresponding thermodynamic bounds.}
\end{figure}

In this section, we discuss how to reach the thermodynamic bounds for proofreading schemes. For convenience, we set $\beta=1$ in the following discussion. In Fig.~\ref{fig:proofreading_SI}a, we illustrate a basic one-step proofreading network, also presented in the main text. At thermodynamic equilibrium, i.e. $\Delta\mu=\ln\frac{k_1k_2k_3}{k_{-1}k_{-2}k_{-3}}=\frac{k'_1k'_2k'_3}{k'_{-1}k'_{-2}k'_{-3}}=0$, there is no proofreading and the two states $EW^*$ and $ER^*$ are discriminated by their energy difference so that the error rate is $\eta_{\varepsilon}=e^{-\epsilon}$. Hopfield showed that better discrimination can be achieved by adding a proofreading step and driving the system out of equilibrium \cite{hopfield1974kinetic}. The Hopfield error rate, $\eta_{h}$, is the square of the equilibrium error rate, i.e., $\eta_{h} = e^{-2\epsilon}$. It is easy to show that the Hopfield error rate can be recovered employing a kinetic symmetry constrain $k_i=k_i'$ for all $i$, and considering the infinitely far-from-equilibrium limit $\Delta\mu\to \infty$, i.e., an infinite amount of available energy. In the most general case, the thermodynamic bounds are shown in Fig.~\ref{fig:proofreading_SI}a and reported in the main text. To saturate these bounds, the kinetic symmetry constraint is loosened to allow a dominating reaction pathway to be kinetically favored. This leads to a significant kinetic discrimination effect when the transition from $E$ and $EW$ is negligible.

The dominating pathway argument can be generalized to analyze the thermodynamic cost of discrimination in networks with arbitrary topology. The most efficient discrimination pathway can be identified by comparing the thermodynamic driving forces along different pathways. One simple example is an N-step proofreading network as illustrated in Fig.~\ref{fig:proofreading_SI}b, which contains $n$ intermediate states before reaching the final releasing states $EW^*$ or $ER^*$. Energy is injected to drive the switching between the intermediate states. The pathways that maximize (minimize) the driving force between the two final states determines the upper (lower) universal thermodynamic bound in Fig.~\ref{fig:proofreading_SI}b. The minimal error is reached by choosing kinetic parameters so that the kinetic rates on the chemical pathways in gray (see Fig.~\ref{fig:proofreading_SI}b) are negligible. This choice allows us to reach the optimal driving force from $EW^*$ to $ER^*$.

\section{Reaction-diffusion pattern}\label{sec:minimal_model}

\subsection{Bounds on visibility of reaction-diffusion pattern}
Reaction-diffusion pattern originate from chemical reactions with bi-stability. For mass-conserving reaction-diffusion systems, a phase-space geometry approach can be applied to study various properties of reaction-diffusion pattern \cite{brauns2020phase}. Our thermodynamic bounds determine the accessible phase-space as a function of the available energy, and thus set limits on such properties. To illustrate this, we consider a simple reaction-diffusion system with two species $U$ and $V$.

\begin{figure}[h]
\centering
\includegraphics[width=\columnwidth]{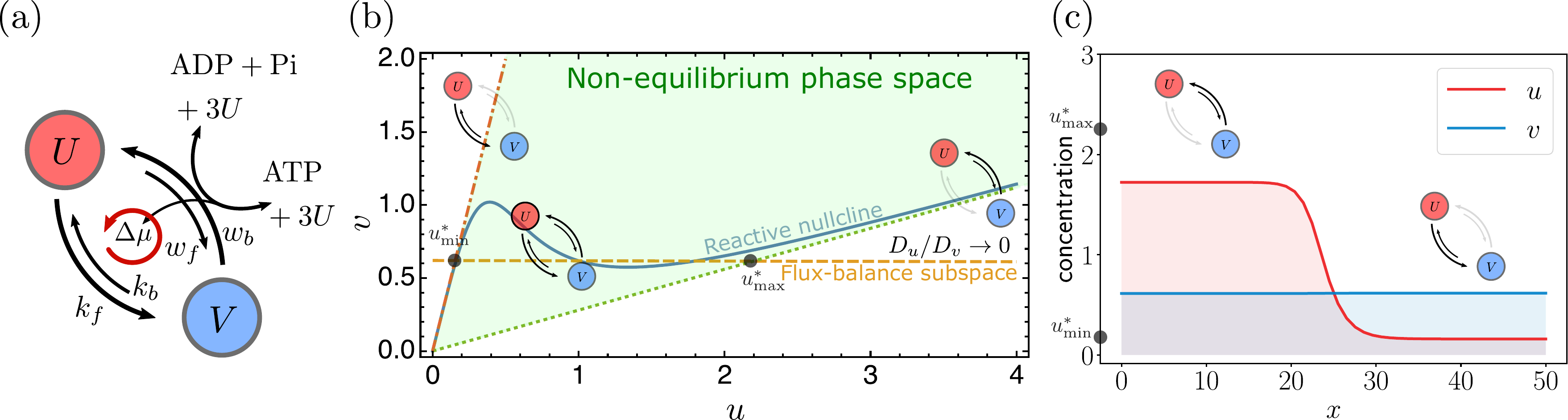}
\caption{\label{fig:RD_SI_1}(a) Active autocatalytic model. (b) The boundaries of the non-equilibrium phase space are determined by two reaction pathways. The autocatalytic reaction scheme shapes the reactive nullclines, enabling switching between these pathways.  (c) Stationary state $u$ and $v$ patterns. The fast-diffusion limit of species $V$ results in a uniform $v$-pattern, while the bi-stable nature of the autocatalytic reaction creates the two plateaus of $u$-pattern.}
\end{figure}

An active (high-order) autocatalytic reaction is one of the simplest chemical reaction models that can show multi-stability. In such a reaction system, we consider two reaction pathways, one describing spontaneous transitions and the other actively fueling the autocatalytic reaction. The reactions are schematized as follows (and shown in Fig.~\ref{fig:RD_SI_1}a):
\begin{equation*}
    \begin{aligned}
        U &\underset{k_b}{\stackrel{k_f}\rightleftharpoons}  V\\   \underbrace{\mathrm{ADP}+\mathrm{Pi}}_\mathrm{Waste}+nU + U&\underset{w_b}{\stackrel{w_f}\rightleftharpoons} V + nU+\underbrace{\mathrm{ATP}}_\mathrm{Fuel}
    \end{aligned}
\end{equation*}
Here the two reaction pathways lead to different equilibrium states. As illustrated in Fig.~\ref{fig:RD_SI_1}b, the presence of both branches allows the system to evolve towards a non-equilibrium stationary state. The non-linear term controls the switch between the reaction branches. When the total concentration is small, the catalytic reaction is negligible and the system mainly stays in the equilibrium state determined by the spontaneous branch. In the other limit, with high concentration, the catalytic branch dominates and the system approaches the (effective) equilibrium state compatible with the energy balance of the coupled ATP hydrolysis. Notice that here we are ignoring the presence of the nucleotide-exchange reactions to avoid complications that will not change the main message.

This system can be easily written in the form of a non-linear rate equation with two distinct reaction pathways between $U$ and $V$. Then, the complete reaction-diffusion equations for the concentrations $u$ and $v$ are
\begin{equation}
    \begin{aligned}
        \frac{\partial u}{\partial t}&=D_u\nabla^2 u+(\hat{k}_{uv}^{(1)}+\hat{k}_{uv}^{(2)})v-(\hat{k}_{vu}^{(1)}+\hat{k}_{vu}^{(2)})u,\\
        \frac{\partial u}{\partial t}&=D_v\nabla^2 v+(\hat{k}_{vu}^{(1)}+\hat{k}_{vu}^{(2)})u-(\hat{k}_{uv}^{(1)}+\hat{k}_{uv}^{(2)})v,\\
    \end{aligned}
\end{equation}
with the effective rates $\hat{k}_{uv}^{(1)}=k_b$, $\hat{k}_{vu}^{(1)}=k_f$, $\hat{k}_{uv}^{(2)}=u^nw_b[\mathrm{ATP}]$ and $k_{vu}^{(2)}=u^nw_f[\mathrm{ADP}][\mathrm{Pi}]$.
These rates satisfy the local detailed balance condition in Eq.~\eqref{eq:local detailed balance}. Here, the linear reaction pathway describes spontaneous transitions and is characterized by the following equilibrium constant:
\begin{equation}
    K_{vu}^\mathrm{max}=K_{vu}^{(1)}\equiv \frac{\hat{k}_{vu}^{(1)}}{\hat{k}_{uv}^{(2)}}=e^{\beta\Delta E_{uv}}
    \label{K1}
\end{equation}
where $\Delta E_{uv}$ is the energy difference between the two states. In contrast, the active autocatalytic branch is driven by ATP hydrolysis with a driving force $\Delta\mu = \Delta\mu_0+k_BT(\ln[\mathrm{ATP}]-\ln[\mathrm{ADP}]-\ln[\mathrm{Pi}])$, thus its equilibrium constant is:
\begin{equation}
    K_{vu}^\mathrm{min} = K_{vu}^{(2)}\equiv\frac{\hat{k}_{vu}^{(2)}}{\hat{k}_{uv}^{(2)}} = e^{\beta(\Delta E_{uv}-\Delta\mu)}
    \label{K2}
\end{equation}
The system is globally out of equilibrium due to the presence of a non-zero thermodynamic affinity along the cycle involving both reaction branches, that is $\mathcal{A} = \ln(K_{vu}^{(1)}/K_{vu}^{(2)}) = \beta\Delta\mu$. Clearly, $\mathcal{A}$ is proportional to the available energy from ATP hydrolysis, encoded in the driving force $\Delta \mu$. 

Due to the unbalance between the two reaction pathways (Eq.s~\eqref{K1} and \eqref{K2}), when the system is driven out of equilibrium, the non-equilibrium phase space expands in the $(u,v)$ plane (see Fig.~\ref{fig:RD_SI_1}b). Notice that in equilibrium conditions, i.e., $\Delta \mu = 0$, this collapses into a line with a slope dictated by the energy difference between the two states. The concentration-dependent nature of the autocatalytic reaction makes the reactive nullclines to have a non-trivial s-shape in this plane represented by the blue line in Fig.~\ref{fig:RD_SI_1}b. This allows for multiple intersections with the flux-balance subspace (dashed orange line in Fig.~\ref{fig:RD_SI_1}b). They exactly determine the range of possible concentrations in a stationary pattern. Nevertheless, the equilibrium constants associated with each branch can provide a bound for such concentrations that is reached when the flux-balance subspace is horizontal and the reactive nullcline is attached to the boundaries of the non-equilibrium phase space. In particular, the intersections between the flux-balance subspace and the phase space boundaries give the maximum and minimum $u$, respectively $u_\mathrm{max}^*$ and $u_\mathrm{min}^*$. Their ratio is upper bounded by the available energy, i.e., $u_\mathrm{max}^*/u_\mathrm{min}^* \leq e^{\beta\Delta\mu}$, which is saturated only for a horizontal flux-balance subspace. The derivation simply follows from the geometric construction and considering the flux-balance subspace can only have a negative slope as both $D_u$ and $D_v$ have to be positive. The upper bound of the visibility immediately follows:
\begin{equation}
\frac{u_\mathrm{max}-u_\mathrm{min}}{u_\mathrm{max}+u_\mathrm{min}} = \mathcal{C}_u \leq \mathcal{C}_u^{\max} \equiv \frac{u_\mathrm{max}^*-u_\mathrm{min}^*}{u_\mathrm{max}^*+u_\mathrm{min}^*} \leq \tanh\left[\frac{\beta\Delta\mu}{2}\right] \equiv \mathcal{B}_{\rm th},
\label{THbound}
\end{equation}
where $\mathcal{B}_{\rm th}$ is the name for this thermodynamic bound. To verify this bound, we did numerical simulations of 1D diffusion-reaction system and plotted the results in the Fig.4c of the main text. In the simulation, we use the 3-rd order autocatalytic reaction with the parameters $k_b+k_f=0.1$, $w_f[\mathrm{ADP}][\mathrm{Pi}]+w_b[\mathrm{ATP}]=1$, $K_{vu}^{\rm max}=4$, $D_u=0.05$, $D_v=25$ and system size $L=50$. The pattern in Fig.~\ref{fig:RD_SI_1}c is obtained with the same parameters and $\beta\Delta\mu = 2.6$.

\begin{figure}[t]
    \centering
    \includegraphics[width=\columnwidth]{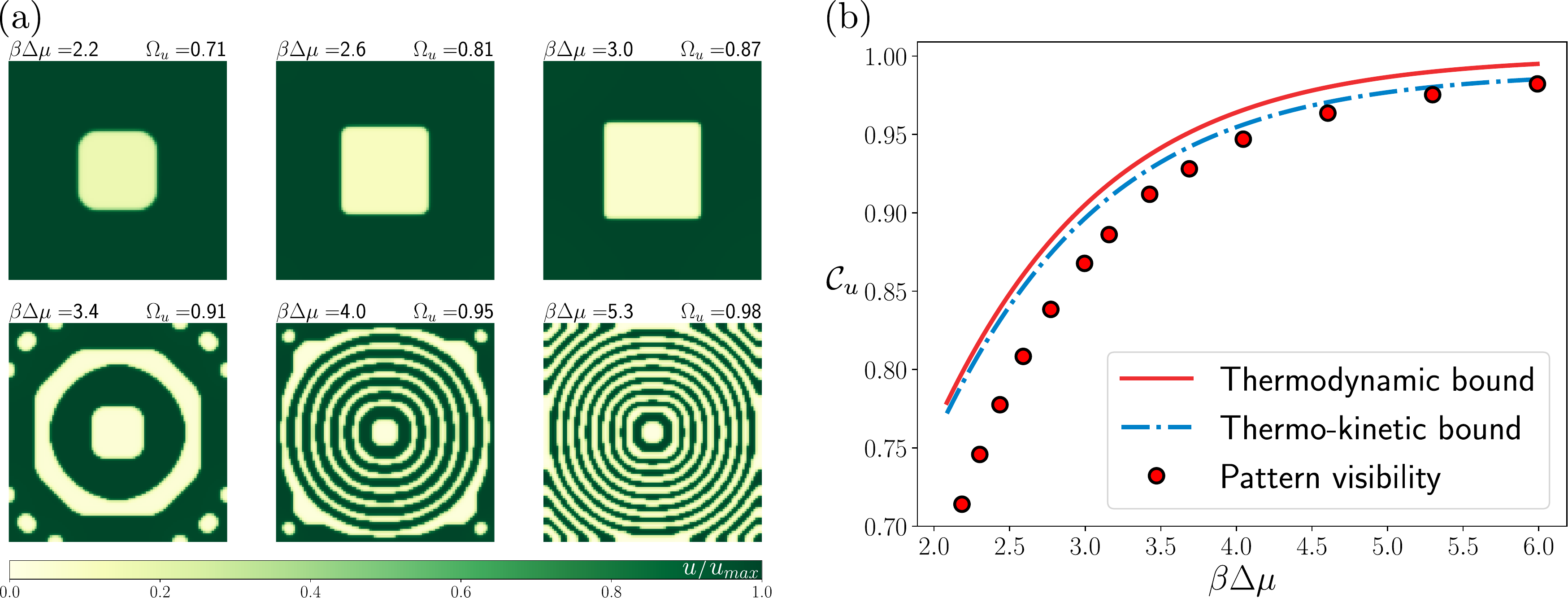}
    \caption{\label{fig:RD_2D}(a) 2D reaction-diffusion patterns under varying driving forces. (b) Thermodynamic and thermo-kinetic bounds on the visibility of the 2D pattern. Adding information about the kinetics clearly improves the estimate. Here, $D_u = 0.1$, $D_v = 5$, systems size $100 \times 100$, and all the other kinetic parameters are the same as in Fig.~\ref{fig:RD_SI_1}.}
\end{figure}

\subsection{Thermo-kinetic tighter bound}
The thermodynamic bound is universal and independent of the kinetics of the system. However, we can get a tighter bound, which we call the thermo-kinetic bound, by including information about the kinetics. The bound on contrast comes from the range of concentration defined by the intersections between the flux-balance subspace, $\eta_0 = D_u u+D_v v$, and the two boundaries of the non-equilibrium phase space, $u/v=K_{vu}^\mathrm{max/min}$. Instead of providing an upper limit for the ratio $u_\mathrm{max}^*/u_\mathrm{min}^*$, we can explicitly solve these intersections as follows:
\begin{equation}
    \begin{aligned}
u_\mathrm{max}^* = \frac{\tilde{\eta}}{R_D + K_{vu}^\mathrm{min} },\quad u_\mathrm{min}^* = \frac{\tilde{\eta}}{R_D + K_{vu}^\mathrm{max}},
\end{aligned}
\end{equation}
where $R_{D}=D_u/D_v$ and $\tilde{\eta}=\eta_0/D_v$. The solution give a thermo-kinetic bound on pattern contrast as
\begin{equation}
    \begin{aligned}
    \mathcal{C}_{u}^{\max} = \tanh\left[\frac{\beta\Delta\mu - \gamma_{u}}{2}\right] \equiv \mathcal{B}_{\rm th-kin},
    \end{aligned}
\end{equation}
where $\gamma_u = \ln\left[\frac{R_D/K_{vu}^{\min}+1}{R_D/K_{vu}^{\max}+1}\right]$. It is evident that $\gamma_u$ is positive and approaches zero in $R_D\to 0$ limit. In Fig.~\ref{fig:RD_2D}a, we show several $2D$ patterns with different values of model parameters and in Fig.~\ref{fig:RD_2D}b, we show that the thermo-kinetic bound provides a better estimate of the actual visibility of reaction-diffusion patterns. Therefore the thermodynamic bound can be recovered when the diffusion of $V$ is infinitely faster than the diffusion of $U$:
\begin{equation}
        \lim_{\frac{D_u}{D_v}\to 0} \mathcal{C}_u^{\max}=\tanh\left[\frac{\beta\Delta\mu}{2}\right].
\end{equation}
This limit agrees with our previous analysis that the maximum possible contrast of a pattern is saturated when the other species is in a uniform distribution due to fast diffusion. Analogously, Eq.~\eqref{THbound} is saturated and the thermodynamic bound, $\mathcal{B}_{\rm th}$, coincides with $\mathcal{B}_{\rm th-kin}$ when the flux-balance subspace is a horizontal line.

\section{Bounds for Chemical Master Equation}\label{sec:CME}

Every chemical system can be written in the form of a chemical master equation (CME). The state of a CME is defined by the number of molecules of each chemical species. For example, if we have $N$ chemical species, a generic state is $\vec{n} = \{n_1, n_2, \dots, n_N\}$ with constraints coming from the fixed number of molecules per species. For example, if only the total number of molecules is fixed to $M_{tot}$, then $\sum_i n_i = M_{tot}$. Conversely, if $A$ and $B$ can form a complex but cannot convert one into another, two constraints have to be put on the total number of molecules in both species. At any rate, we have a CME defined in terms of the probability to be in a given state $\vec{n}$, i.e., $p(\vec{n})$, plus the normalization condition. A CME cannot always be written as a rate equation with only catalytic reactions, e.g., when complexes are involved and when we cannot take the limit of large number of molecules. However, for non-complex species, it is reasonable to define the average probability to observe a given state, $k$, as follows:
\begin{equation}
    \mathcal{P}(k) = \frac{\langle n_k \rangle}{M_{tot}}
\end{equation}
In the absence of complexes, $\mathcal{P}(k)$ satisfies a rate equation in the limit of a very high number of molecules, otherwise fluctuations play an important role. For a multi-molecular complex, the definition of an analogous quantity involves the use a normalization factor at the denominator that might be different from $M_{tot}$, depending on whether or not the interest is to quantity the average amount of complexes with respect to total monomers.

\subsection{Bounds on the ratio of averages}
We can define a symmetry-breaking index as we did before:
\begin{equation}
    s = \frac{\mathcal{P}(i_1)}{\mathcal{P}(j_1)} = \frac{\sum_{\vec{n}} n_{i_1} p(\vec{n})}{\sum_{\vec{n}} n_{j_1} p(\vec{n})} = \frac{\sum_{n_{i_1}} n_{i_1} \sum_{\vec{n}/n_{i_1}} p(\vec{n})}{\sum_{n_{j_1}} n_{j_1} \sum_{\vec{n}/n_{j_1}} p(\vec{n})}
\end{equation}
where the notation $\vec{n}/n_{i_1}$ indicates that the sum is performed over all indices except for $n_{i_1}$. Notice that the space of possible states is way larger than before, since we are tracking the number of molecules in each chemical species.

Since we are now dealing with monomers in quantifying the symmetry breaking, both $n_{i_1}$ and $n_{j_1}$ can be at least $0$ and at most $M_{tot}$. First, we define $\mathbf{i}_m = (i_1, \dots, i_m)$ as an $m$-dimensional vector whose values range from $1$ to $N$, indicating the species, and $\mathbf{a}_m = (a_1, \dots, a_m)$ as an $m$-dimensional vector whose elements span all possible number of molecules (per species) compatible with the chemical space. Thus, we introduce the following short notation:
\begin{equation*}
\mathbf{n}_{\mathbf{i}_1}^{\mathbf{a}_1} = p(n_1, \dots, n_{i_1} = a_1, \dots, n_N) \qquad a_1 = {0, \dots, M_{tot}}
\end{equation*}
This is a coarse-grained state accounting for all possible states with $n_{i_1} = a_1$. In this particular case, $a_1$ can take $M_{tot} + 1$ values, from $0$ to $M_{tot}$. The symmetry-breaking index reads:
\begin{equation}
s = \frac{\sum_{a_1} a_1 \sum_{\vec{n}/n_{i_1}}^{i_1,a_1} \mathbf{n}_{\mathbf{i}_1}^{a_1}}{\sum_{a_1} a_1 \sum_{\vec{n}/n_{j_1}}^{j_1,a_1} \mathbf{n}_{\mathbf{j}_1}^{a_1}} = \frac{1 \sum_{\vec{n}/n_{i_1}}^{i_1,1} \mathbf{n}_{\mathbf{i}_1}^{a_1 = 1} + 2 \sum_{\vec{n}/n_{i_1}}^{i_1,2} \mathbf{n}_{\mathbf{i}_1}^{a_1 = 2} + \dots}{1 \sum_{\vec{n}/n_{j_1}}^{j_1,1} \mathbf{n}_{\mathbf{j}_1}^{a_1 = 1} + 2 \sum_{\vec{n}/n_{j_1}}^{j_1,2} \mathbf{n}_{\mathbf{j}_1}^{a_1 = 2} + \dots} \;,
\end{equation}
where $\sum_{\vec{n}/n_{i_1}}^{i_1,a_1}$ indicates a summation over all $\vec{n}$ but $n_{i_1}$, which is fixed to $\mathbf{a}_1$ instead. Moreover, since $p(\vec{n})$ is governed by a Master Equation, its steady-state solution can be expressed in terms of spanning trees, as before. Thus, we have:
\begin{equation}
s = \frac{\sum_{a_1} a_1 \sum_{\vec{n}/n_{i_1}}^{i_1,a_1} \sum_{\mu} A_{i_1}^{a_1}(T_\mu)}{\sum_{a_1} a_1 \sum_{\vec{n}/n_{j_1}}^{j_1,a_1} \sum_{\mu} A_{j_1}^{a_1}(T_\mu)} \;,
\end{equation}
where $A_{i_1}^{a_1}(T_\mu)$ is a short notation indicating the product of all the rates belonging to the spanning tree $T_\mu$, oriented towards the state $\vec{n}$ where $n_{i_1} = a_1$. Applying the inequality shown above by maximizing only over $T_\mu$, we obtain:
\begin{equation}
s \leq \max_{\{ T_{\mu} \}} \left( \frac{\sum_{\vec{n}} a_1 K^{\rm eq}_{(i_1,a_1),(i_1,L)}(T_\mu)}{\sum_{\vec{n}} a_1 K^{\rm eq}_{(j_1,a_1),(j_1,L)}(T_\mu)} K^{\rm eq}_{(i_1,L),(j_1,L)}(T_\mu) \right),
\end{equation}
by indicating with $K^{\rm eq}_{(i_1,a_1),(i_1,L)}(T_\mu) = A_{i_1}^{a_1}(T_\mu)/A_{i_1}^{L}(T_\mu)$ and selecting $a_1 = L$ as a reference state for both $i_1$ and $j_1$ for simplicity. The derivation follows the same steps shown above to derive the universal thermodynamic bounds for a set of states. Clearly, we can maximize also over $a_1$, obtaining:
\begin{equation}
s \leq \max_{\{ T_{\mu} \}, a_1} \left( \frac{\sum_{\vec{n}/n_{i_1}} K^{\rm eq}_{(i_1,a_1),(i_1,L)}(T_\mu)}{\sum_{\vec{n}/n_{j_1}} K^{\rm eq}_{(j_1,a_1),(j_1,L)}(T_\mu)} K^{\rm eq}_{(i_1,L),(j_1,L)}(T_\mu) \right),
\end{equation}
Finally, by maximizing also over all the remaining $\vec{n}$, we obtain the following (loose) bound:
\begin{equation}
s \leq \max_{\{ T_{\mu} \}, \{ \vec{n} \}, a_1} \left( K^{\rm eq}_{(i_1,a_1),(j_1,a_1)}(T_\mu) \right),
\end{equation}
When extending these bounds to the case in which we evaluate ratios between multi-molecular complexes and monomers, we note that the sum on the numerator and denominator runs over different indices, since complexes cannot be populated up to $M_{tot}$ and they also might not populate the entire range of values. If the species $c$ is the complex, and $n_c \in N_c$, with $N_c$ the subset of accessible states, then $N_c > 0$ and $\max(N_c) = M_c \leq M_{tot}$. Thus:
\begin{equation*}
s = \frac{\langle n_i \rangle}{\langle n_c \rangle} = \frac{\sum_{n_i=0}^{M_{tot}} n_i \sum_{\vec{n}/n_i} p(\vec{n})}{\sum_{n_c \in N_c} n_c \sum_{\vec{n}/n_c} p(\vec{n})} \;.
\end{equation*}
Starting from this expression, we can maximize over $T_\mu$. If we want to provide looser bounds that use less information and maximize also over the value that $n_i$ and $n_c$ can take, we need to restrict the summation at the numerator:
\begin{equation*}
s = \frac{\langle n_i \rangle}{\langle n_c \rangle} = \frac{\sum_{n_i=0}^{M_{tot}} n_i \sum_{\vec{n}/n_i} p(\vec{n})}{\sum_{n_c \in N_c} n_c \sum_{\vec{n}/n_c} p(\vec{n})} \leq \frac{\sum_{n_i \in N_c} n_i \sum_{\vec{n}/n_i} p(\vec{n})}{\sum_{n_c \in N_c} n_c \sum_{\vec{n}/n_c} p(\vec{n})} \;,
\end{equation*}
and everything follows as before. If this restriction cannot be performed - when the two summation runs on disjoint sets of indices as for the ratios of two complexes - the maximization can be performed only over $T_\mu$. Notice that the bounds obtained for a CME are looser than those for rate equations, even if they apply to all possible scenarios.

\subsection{Bounds on the ratio of correlations}
Another advantage of the CME is that the universal thermodynamic bounds can be derived for ratios of any correlation. The symmetry-breaking index can be defined as follows:
\begin{equation}
s_l = \frac{\langle n_{i_1} n_{i_2} \dots n_{i_l} \rangle}{\langle n_{j_1} n_{j_2} \dots n_{j_l} \rangle} = \frac{\sum_{\vec{n}} \mathbf{a}_l ~\mathbf{n}^{\mathbf{a}_l}_{\mathbf{i}_l}}{\sum_{\vec{n}} \mathbf{a}_l ~\mathbf{n}^{\mathbf{a}_l}_{\mathbf{j}_l}} \leq \max_{\{T_\mu\}} \left( \frac{\sum_{\vec{n}} \mathbf{a}_l ~K^{\rm eq}_{(\mathbf{i}_l,\mathbf{a}_l),(\mathbf{i}_l,\mathbf{L})}}{\sum_{\vec{n}} \mathbf{a}_l ~K^{\rm eq}_{(\mathbf{j}_l,\mathbf{a}_l),(\mathbf{j}_l,\mathbf{L})}} K^{\rm eq}_{(\mathbf{i}_l,\mathbf{L}),(\mathbf{j}_l,\mathbf{L})} \right)
\end{equation}
Notice that now $\mathbf{a}_l$ and $\mathbf{i}_l$ are $l$-dimensional vectors. Here, we used $\mathbf{i}_l = \mathbf{L}$ and $\mathbf{j}_l = \mathbf{L}$ as reference states for simplicity. This limit can be upper bounded (maximizing over all variables) as follows:
\begin{equation}
s_l \leq \max_{\{T_\mu\}, \{\vec{n}\}, \mathbf{a}_l} \left( K^{\rm eq}_{(\mathbf{i}_l,\mathbf{a}_l),(\mathbf{j}_l,\mathbf{a}_l)} \right)
\end{equation}

\bibliography{refs}